\documentstyle[twocolumn,aps,psfig]{revtex}

\begin{document}
\title{ Geometric frustration in the mixed layer pnictide oxides }
\author{Matthew Enjalran }
\address{Department of Physics, University of California, Davis, CA 95616 \\
and Materials Research Institute, Lawrence Livermore National
Laboratory, \\
University of California, Livermore, CA 94550}
\author{Richard T. Scalettar}
\address{Department of Physics, University of California, Davis, CA 95616}
\author{Susan M. Kauzlarich}
\address{Department of Chemistry, University of California, Davis, CA 95616}
\date{\today}
\maketitle

\begin{abstract}

We present results from a Monte Carlo investigation of 
a simple bilayer model with geometrically frustrated interactions
similar to those found in the mixed layer 
pnictide oxides $(Sr_{2}Mn_{3}Pn_{2}O_{2}, Pn=As,Sb).$  Our model
is composed of two inequivalent square lattices with nearest neighbor
intra- and interlayer interactions. We find a ground state composed of 
two independent N\'{e}el ordered layers when the interlayer exchange
is an order of magnitude weaker than the intralayer exchange, as suggested by
experiment. We observe this result independent of the number of 
layers in our model. We find evidence for local orthogonal order
between the layers, but it occurs in regions of parameter space that are not
experimentally realized. We conclude that frustration caused by
nearest neighbor interactions in the mixed layer pnictide oxides 
is not sufficient to explain the long--range orthogonal order that is 
observed experimentally, and that it is likely that other terms (e.g.,
local anisotropies) in the Hamiltonian are required to explain the
magnetic behavior.  
\end{abstract}


\section{Introduction}
\label{sect-intro}

Clean systems of interacting moments 
have been studied extensively by analytic and numerical techniques.  
Although simplified models like Ising, Heisenberg, and Hubbard  
retain only the most fundamental interactions observed in real materials,
they remain tractable to current theoretical techniques, and 
the study of their ordered phases in various regions of parameter
space has contributed enormously to our understanding of magnetic 
phenomena and the physics of correlated systems.~\cite{magn-books}  
However, real materials are never clean. There is often 
frustration due to competing interactions and 
disorder in the interaction strengths. 

Competing interactions that cause magnetic frustration can have many 
origins, lattice geometry, magnetic and non-magnetic impurities. 
In three dimensions, helical magnetic order
has been observed when geometric frustration is accompanied by
anisotropy.~\cite{plumer1,diep} Spin glasses phases are
observed when frustration is accompanied by random 
disorder.~\cite{binder,young,gingras}
It has also been suggested that some
non-collinear spin ordered structures belong to a 
new chiral universality class.~\cite{kawamura} 
The systems we study are essentially two
dimensional and contain no
anisotropic terms or disorder. Frustration is caused by the lattice
geometry. Our primary focus is the orthogonal magnetic 
structure observed in the mixed layer pnictide oxides.~\cite{kauz}  

The pnictide oxides of type $A_{2}Mn_{3}Pn_{2}O_{2} 
(A = Ba, Sr; Pn = As, Sb)$
are layered antiferromagnets that contain   
two distinct square planes of manganese atoms arranged in a lattice 
of space group symmetry $I4/mmm.$ In one layer, 
manganese is bonded to oxygen
in a planar $CuO_{2}$ arrangement, $MnO_{2}^{2-}.$  
In a second layer, it is bonded to a pnictogen in
a tetrahedral structure, $MnPn_{2}^{2-},$ where pnictogen atoms
project alternately above and below the plane defined by the 
manganese atoms. From here on we denote the two layers as 
Mn(1) for $MnO_{2}^{2-}$ and Mn(2) for $MnPn_{2}^{2-}.$ The
Mn atoms from the two planes are arranged so that a
site in the Mn(1) layer sits directly above and below the 
center of a square plaquette of Mn atoms in the 
Mn(2) layer. The manganese carry a spin $S = 5/2.$  
Frustration can enter through nearest neighbor interlayer coupling.
A more detailed investigation of these systems 
has been reported elsewhere.~\cite{enjalran} 

\section{Model} 
\label{sect-model}

In all the pnictide oxides except 
$Sr_{2}Mn_{3}As_{2}O_{2}$ there is long
range order in the planes that eventually gives rise to 
weak $3D$ order. In the compound $Sr_{2}Mn_{3}As_{2}O_{2},$
there is only short range order in the Mn(1) planes. 
The ability of ordered planes to drive
c-axis order has been investigate before in the case of 
layered antiferromagnets.~\cite{lines,birgeneau,birgeneau2} 
In $Sr_{2}Mn_{3}Sb_{2}O_{2}$,
magnetic order in the Mn(1) layers is established along 
the a-axis of the magnetic unit cell, while in the Mn(2) layers 
the magnetization is along
the c-axis. Hence there is an orthogonal alignment between neighboring
layers. Such an ordered state is not without
precedent.~\cite{leciej,torardi} However, the
different temperatures at which the layers order  
($T_{Mn2} \approx 300K$ and $T_{Mn1} \approx 65K$) and the symmetry of the 
frustrated interlayer interactions which leads to cancellation suggest
a system of two independent N\'{e}el
ordered layers.  Previous work has shown that this is not always the 
case, as thermal or quantum
fluctuations (in frustrated systems) 
can lift the degeneracy of the system to select a single
state.~\cite{henley,moreo} 

\begin{figure}
\center
\leavevmode
\psfig{file=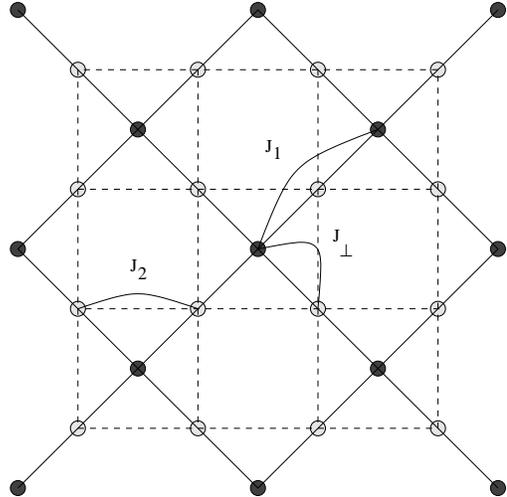,height=2.6in,width=2.6in,angle=-90}
\vspace{1 mm}
\caption {A 2D projection of the two distinct layers of the 
pnictide oxide $Sr_{2}Mn_{3}Sb_{2}O_{2}.$ 
Sites in the Mn(1) layer are represented by dark circles while
sites in the Mn(2) layer are represented by light circles.  
The intralayer couplings are shown as $J_{1}$ and $J_{2}$, 
and the interlayer interaction is indicated by $J_{\perp}$. }
\label{fig-exptgeo}
\end{figure}

To study the effect of
frustration on the ground state 
magnetic order of the pnictide oxides, 
we developed a simple model of classical
Heisenberg spins with nearest neighbor intra- and interlayer 
interactions. The basic structural unit is a set of two layers, 
one each of type Mn(1) and $Mn2$ (see Fig.~\ref{fig-exptgeo}). 
The Mn(2) layer 
has a lattice constant $a=1$ and contains $n^{2}$ sites. The 
Mn(1) layer is larger by a factor $\sqrt{2}$ and is rotated by
$\pi/4$ with respect to the lattice directions of the other 
layer. The Mn(1) layer contains $n^{2}/2 + n + 1$ spins.   
Note that the interlayer coordination is not the same for 
spins on the two layers. A spin on the Mn(1) plane is coupled
to four spins on the Mn(2) plane, and each spin in the Mn(2) plane 
is coupled to only two spins on the Mn(1) plane. 

The Hamiltonian for our bilayer model is written as 
\begin{eqnarray}
H & = & J_{1} \sum_{i,\vec{\delta}_{1}} \vec{S}_{i}^{(1)} \cdot \vec{S}_{i+\vec{\delta}_{1}}^{(1)} 
  + J_{2} \sum_{i,\vec{\delta}_{2}} \vec{S}_{i}^{(2)} \cdot 
\vec{S}_{i+\vec{\delta}_{2}}^{(2)}  \nonumber \\
  & + & J_{\perp} \sum_{i,\vec{\delta}_{\perp}} \vec{S}_{i}^{(\alpha)} 
\cdot \vec{S}_{i+\vec{\delta}_{\perp}}^{(\beta)}.
\label{eq-ham} 
\end{eqnarray}
The constants $J_{1}, J_{2},$ and $J_{\perp}$ represent the Mn(1) and
Mn(2) intralayer couplings and the interlayer coupling, respectively. 
The summations of $\vec{\delta}_{\mu}$ are over nearest neighbors to
site $i.$  For classical spins, one has 
$|\vec{S}| = (S^{2}_{x} + S^{2}_{y} + S^{2}_{z})^{1/2} =
1$. The relatively large spin-$5/2$ of the Mn atoms in the pnictide
oxides makes this a reasonable approximation.

We studied the equilibrium physics of our model by a 
single spin flip Monte Carlo algorithm. We have addressed concerns
about proper sampling of phase space by performing simulations with
random and ordered initial configurations. We have also considered
the effects of the boundary on our finite simulations 
by employing a few different boundary conditions: open, periodic, and 
periodic with an effective field on the Mn(1) edge sites. 
In all cases considered, we found no qualitative difference in our 
results due to the initial configuration or the 
conditions imposed at the boundary. 

To determine the relative orientation between neighboring spins,
either within the same layer or in different layers, 
we measured a collinear 
\begin{equation}
C_{\parallel}^{\alpha, \beta} = \left \langle \frac{1}{zN_{\alpha}}\sum_{i}\sum_{\vec{\delta}}
(\vec{S}^{\alpha}_{i} \cdot \vec{S}^{\beta}_{i+\vec{\delta}})^{2}
\right \rangle,
\label{eq-collin} 
\end{equation}
and a perpendicular 
\begin{equation}
 C_{\perp}^{\alpha, \beta} = \left \langle \frac{1}{zN_{\alpha}}\sum_{i}\sum_{\vec{\delta}}
(\vec{S}^{\alpha}_{i} \times \vec{S}^{\beta}_{i+\vec{\delta}})^{2}
\right \rangle
\label{eq-perp}
\end{equation}
spin-spin correlation function. Here summations are performed over all
nearest neighbors $\vec{\delta}$ of site i and then over all sites
in the lattice; $z$ is the coordination number and $N_{\alpha}$ is the
number of sites in layer $\alpha.$ Intralayer correlations are denoted by 
$\alpha = \beta$ and interlayer correlations are represented by 
$\alpha \neq \beta.$  We stress that $C_{\parallel}$ and $C_{\perp}$
measure local correlations.  For classical Heisenberg spins, these
correlations take on the simple forms $C_{\parallel} = \langle \cos^{2} \theta \rangle$
and $C_{\perp} = \langle \sin^{2} \theta \rangle.$ In the high
temperature, paramagnetic, limit, the values $C_{\parallel}=1/3$
and $C_{\perp}=2/3$ are obtained. We also measured the magnetization 
and staggered magnetization of each layer. 

\section{Results and Discussion} 
\label{sect-expt-mc}

From the experimental data, representative couplings would set
the Mn(2) intralayer exchange to be stronger than the Mn(1) intralayer
exchange, with the interlayer interaction weaker by at least an order
of magnitude. Therefore, experimentally motivated couplings in our
model were set to $J_{2} = 2.0, J_{1} = 1.0$ and $J_{\perp} = 0.1.$
For the results presented herein, systems with $1600$ spins per
Mn(1) layer and $841$ spins per Mn(1) were equilibrated for 
$15,000$ to $25,000$ sweeps followed by $15,000$ measurement sweeps
with $10-25$ sweeps between measurements. One Monte Carlo sweep
denotes an update of all spins on the lattice. 
 
\begin{figure}
\center
\leavevmode
\psfig{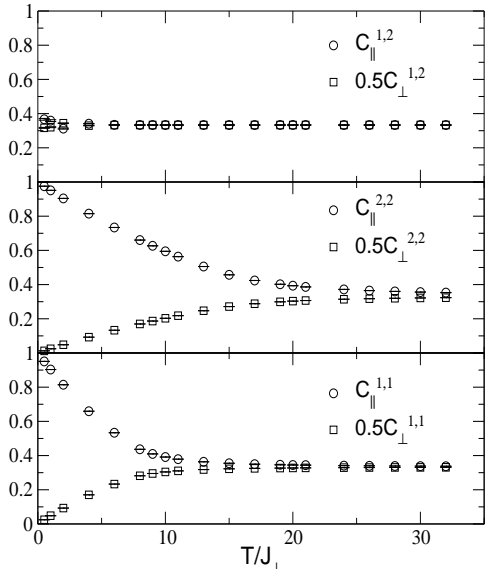}
\caption {Temperature dependence of the local intra- and interlayer 
spin-spin correlations in the four layer model with 
periodic boundary conditions and $J_{1}=1.0, J_{2}=2.0, 
J_{\perp}=0.1.$ A parallel alignment is favored for
intralayer spins when the temperature drops below the respective 
intralayer coupling; however, the interlayer correlations remain at
the high temperature limit of $1/3$ even for 
$T\stackrel{<}{\sim}J_{\perp}.$}
\label{fig-exptcpp}
\end{figure}

Our results for a bilayer model indicate that, as a function of 
temperature, the moments within each layer began to order when the 
temperature dropped below the respective energy scale, e.g., $T=J_{1}$
for Mn(1) moments
and $T=J_{2}$ for Mn(2) moments. However, the eventual ground state
was
a system of
two N\'{e}el ordered layers with an arbitrary orientation between 
the magnetization directions.
In a simulation with four layers (i.e., a layered sequence 
$Mn1-Mn2-Mn1-Mn2$) and periodic boundary conditions along the
c-axis, we observed the same qualitative behavior as a function of 
temperature (see Fig.~\ref{fig-exptcpp}).  
We emphasize that the interlayer spin-spin correlation function,
$C_{\parallel}^{1,2},$ remained at the paramagnetic limit down to 
low temperatures, $T\stackrel{<}{\sim}J_{\perp}.$

\begin{figure}
\center
\leavevmode
\psfig{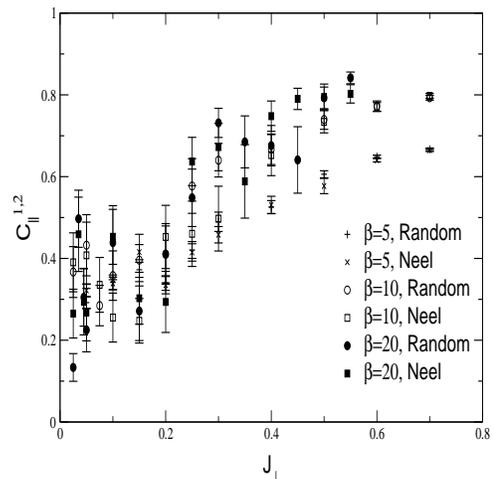}
\caption {Interlayer spin--spin correlations as a function of $J_{\perp}$
with $J_{1}=1.0$ and $J_{2}=2.0.$ The results are for a bilayer model
with the initial configuration of the
layers being either random or N\'{e}el ordered.}
\label{fig-j11j22}
\end{figure}

We also studied the effect of the strength of frustration on the 
magnetic ground state. To do this, we fixed the temperature and 
swept in values of $J_{\perp}.$ For a bilayer model with $J_{\perp} < 0.25,$
we observed two N\'{e}el ordered layers with a paramagnetic 
interlayer orientation, i.e., $C_{\parallel}^{1,2} \approx 1/3$
or independent layers. As $J_{\perp}$ was increased, the two
N\'{e}el ordered layers moved towards a collinear state.  
We observed this behavior independent of the 
initial configuration (see Fig.~\ref{fig-j11j22}). 
 
An explanation of these results can be understood by considering 
the physics of two simpler models, a zigzag lattice and a diagonal lattice, 
(refer to Ref.~\cite{enjalran}). In the zigzag lattice the frustration is along
one crystalline direction; this geometry is similar to the Mn(2) to
Mn(1) interlayer interaction.  In the limit of weak $J_{\perp},$
a uniformly canted state is established in the zigzag lattice 
with orthogonal order obtained in the limit $J_{\perp} \rightarrow 0.$ 
In the diagonal lattice, one has the equivalent of the $J_{1}-J_{2}$
model. This interlayer geometry is analogous to the coupling of
Mn(1) sites to Mn(2) sites.  In the limit of weak $J_{\perp},$ 
the diagonal lattice is composed of two N\'{e}el ordered layers with
a collinear alignment. Hence, at weak frustration, it is the
competition between these two tendencies in
the experimental model that leads to a paramagnetic orientation with 
large fluctuations. In the case of strong frustration, 
$J_{\perp} > 0.5,$ the zigzag and diagonal geometries both select a 
collinear interlayer arrangement. 
 
An orthogonal state for our bilayer model can be found but this phase 
occurs at couplings strengths that are not supported by experiments 
(refer to Ref.~\cite{enjalran}). 
In the case  
where $J_{2}=0,$ the resultant model is a network of intersecting 
zigzag chains.
By setting $J_{\perp}=1$ and sweeping in $J_{1}$ a 
transition to a uniformly canted state was observed in 
the limit of large $J_{1}$.
For a model with $J_{2}=J_{\perp}=1.0,$ we observed a N\'{e}el ordered Mn(2) layer 
and a paramagnetic Mn(1) layer, but with Mn(1) spins 
orthogonal to the local Mn(2) environment, for 
$J_{1} \leq 0.25$. At $J_{1} > 0.25$ the system tended toward a 
state with collinear alignment. 

We conclude that frustration caused by nearest neighbor interactions, 
both intra- and interlayer, in the mixed layer pnictide oxides 
is not sufficient to explain the long range orthogonal order that
is observed experimentally. However, recent work on mixed metal pnictide oxides
does indicate that the two distinct layers do not behave
independently.~\cite{tadashi} In these systems, it is likely
that other terms in the Hamiltonian, e.g., ion anisotropies arising
from spin-orbit effects, are required to explain the magnetic behavior. 
  
We acknowledge the generous support of Campus--Laboratory Collaboration of the 
University of California, the Materials Research Institute at
Lawrence Livermore National Laboratory. 
We thank Rajiv Singh and Tadashi Ozawa at the University of California,
Davis for many useful discussions.  
We also thank D.P. Landau at the Center for Simulational Physics at
the University of Georgia, Athens for helpful suggestions.  
Work at Lawrence Livermore National Laboratory performed under the auspices of the
U.S. Department of Energy under contract number W-7405-ENG-48.



\begin{thebibliography}{99}

\bibitem{magn-books}
D.C. Mattis, \textit{The Theory of Magnetism II, Thermodynamics and
Statistical Mechanics}, edited by P. Fulde, (Springer--Verlag,
Berlin, 1985); A. Auerbach, \textit{Interacting Electrons and Quantum 
Magnetism}, edited by J.L. Birman, H. Faissner, and J.W. Lynn, 
(Springer--Verlag, New York, 1994); F. Gebhard, \textit{The Mott 
Metal--Insulator Transition}, edited by G. H\"{o}hler,
(Springer--Verlag, Berlin, 1997).
 

\bibitem{plumer1}
M.L. Plumer, A. Caill\'{e}, and H.T. Diep, in
\textit{Magnetic Systems with Competing
Interactions (Frustrated Spin Systems)}, edited by H.T. Diep, 
(World Scientific, 1994).

\bibitem{diep}
H.T. Diep, Europhys. Lett. \textbf{7}, 725 (1988);
H.T. Diep, Phys. Rev. B \textbf{39}, 397 (1989).
 
\bibitem{binder}
K. Binder and A.P. Young, Rev. Mod. Phys. \textbf{58,} 801 (1986).

\bibitem{young}
\textit{Spin Glasses and Random Fields}, edited by A.P. Young,
(World Scientific, River Edge, N.J., 1998).

\bibitem{gingras}
M.J.P. Gingras, in
\textit{Magnetic Systems with Competing
Interactions (Frustrated Spin Systems)}, edited by H.T. Diep, 
(World Scientific, 1994).

\bibitem{kawamura}
H. Kawamura, J. Phys. Soc. Jpn. \textbf{54}, 3220 (1985);
H. Kawamura, Phys. Rev. B \textbf{38}, 4916 (1988).

\bibitem{kauz} 
S.L. Brock, N.P. Raju, J.E. Greedan, and S.M. Kauzlarich, 
J. Alloys Comp. \textbf{237}, 9 (1996);
S.L. Brock and S.M. Kauzlarich, 
J. Alloys Comp. \textbf{241}, 82 (1996);
S.L. Brock and S.M. Kauzlarich, 
Chemtech \textbf{25}, 18 (1995).

\bibitem{enjalran} 
M. Enjalran, R.T. Scalettar, and S. M. Kauzlarich,
Phys. Rev. B \textbf{61}, 14570 (2000).

\bibitem{lines}
M.E. Lines, Phys. Rev. \textbf{164}, 736 (1967);
M.E. Lines, J. Phys. Chem. Solids \textbf{31}, 101 (1970);

\bibitem{birgeneau}
R.J. Birgeneau, H.J. Guggenheim, and G. Shirane, Phys.
Rev. Lett. \textbf{22}, 720 (1969).

\bibitem{birgeneau2}
R.J. Birgeneau, H.J. Guggenheim, and G. Shirane, Phys.
Rev. B \textbf{1}, 2211 (1970); R.J. Birgeneau, 
H.J. Guggenheim, and G. Shirane, Phys. Rev. B \textbf{8}, 304 (1973). 

\bibitem{leciej}
J. Leciejewicz, S. Siek, and A. Szytula, J. Magn. 
Mater. \textbf{40}, 265 (1984).

\bibitem{torardi}
C.C. Torardi, W.M. Reiff, K. L\'{a}z\'{a}r, J.H. Zhang, and D.E. Cox,
J. Solid State Chem. \textbf{66}, 105 (1987).
 
\bibitem{henley}
C. Henley, Phys. Rev. Lett. \textbf{62}, 2056 (1989).

\bibitem{moreo}
A. Moreo, E. Dagotto, T. Jolicoeur, and J. Riera,
Phys. Rev. B \textbf{42}, 6283 (1990).

\bibitem{tadashi}
T. Ozawa, M.M. Olmstead, S.L. Brock, S.M. Kauzlarich,
and D.M. Young, Chem.\ Mater.\ \textbf{10}, 392 (1998);
T. Ozawa, S.M. Kauzlarich, M. Bieringer, C.R. Wiebe, J.E.
Greedan, and J.S. Gardner, submitted to Chem.\ Mater.

\end{thebibliography}
\end{document}